\documentclass[english,showkeys,notitlepage]{revtex4}
\usepackage[T1]{fontenc}
\usepackage[latin9]{inputenc}
\usepackage{babel}
\usepackage{amsmath}
\usepackage{amssymb}
\usepackage{graphicx}
\usepackage{esint}
\usepackage[unicode=true,pdfusetitle,
 bookmarks=true,bookmarksnumbered=false,bookmarksopen=false,
 breaklinks=false,pdfborder={0 0 1},backref=false,colorlinks=false]
 {hyperref}

\makeatletter
\@ifundefined{textcolor}{}
{%
 \definecolor{BLACK}{gray}{0}
 \definecolor{WHITE}{gray}{1}
 \definecolor{RED}{rgb}{1,0,0}
 \definecolor{GREEN}{rgb}{0,1,0}
 \definecolor{BLUE}{rgb}{0,0,1}
 \definecolor{CYAN}{cmyk}{1,0,0,0}
 \definecolor{MAGENTA}{cmyk}{0,1,0,0}
 \definecolor{YELLOW}{cmyk}{0,0,1,0}
}

\usepackage{babel}

\usepackage{babel}

\usepackage{babel}
\usepackage{babel}

\usepackage{babel}

\makeatother

\begin{document}

\title{Strangeness production in high-multiplicity events}

\author{ Marat Siddikov, Iván Schmidt}

\affiliation{Departamento de Física, Universidad Técnica Federico Santa María,~~~\\
 y Centro Científico - Tecnológico de Valparaíso, Casilla 110-V, Valparaíso,
Chile}
\begin{abstract}
In this paper we analyze in detail the production of strangeness in
proton-proton collisions in the kinematics of large transverse momenta
$p_{T}$ of produced hadrons. Using the color dipole framework, we estimated the production
cross-sections for kaons and demonstrated 
that the shapes of the $p_{T}$-dependence are in agreement with available
experimental data. We also analyzed the self-normalized yields of
strange hadrons as a function of multiplicity of co-produced hadrons,
and found that the predictions are in agreement with the faster-than-linear
growth seen in experimental data. Our description is largely parameter-free
and complements our previous studies dedicated to the explanation
of multiplicity enhancement of quarkonia, as well as $D$- and $B$-mesons. 
\end{abstract}

\date{\today}

\maketitle

\section{Introduction}

Since the early experiments at RHIC and SPS~\cite{SPS1,SPS2,SPS3,SPS4,SPS5,SPS6},
the production of hadrons containing strange quarks has been used
as one of the probes of Quark-Gluon Plasma (QGP) formation, described
in the framework of QGP-inspired hydrodynamic models~\cite{Rafelski:1982ab,Rafelski:1982ck,Koch:1982df,Letessier:1993,Letessier:1994,Rafelski:1994gh,Sollfrank:1994,Letessier:1995,BraunMunzinger:1995,Harris:1996zx,Koch:2017pda,Rafelski:1982pu,Rafelski:2019twp,Kopeliovich:2017jpy,Schmidt:2018rkw}.
The QGP manifests itself in different ways in experimental observables.
For example, it might lead to an enhancement of strange particle yields
in heavy ion collisions, compared to $pA$ and $pp$ production of
the same strange hadrons. Another possibility to observe the effects
of QGP is via the enhancement of strangeness in events with large
multiplicity of co-produced hadrons. This observable might be studied
independently in $pp$, $pA$ or $AA$ collisions. While early experiments
confirmed the enhancement of strangeness in heavy ion collisions,
similar multiplicity enhancements has been recently observed at the
LHC not only in heavy ion, but also in $pA$~\cite{ALICE:pAStrangeness,ALICE:pAStrangeness2}
and even in $pp$ collisions~\cite{ALICE:2017jyt}, where QGP formation
in significant amounts is highly unlikely even at TeV-range collision
energies. For this reason it makes sense to understand better the
microscopic mechanisms of this phenomenon, at least in $pp$ collisions.
In general, application of perturbation theory for strangeness production
is challenging due to lack of the hard scale (like e.g. heavy mass
of the quark). Threfore the phenomenological description of strangeness
production has been mostly limited to studies in the framework of
Monte-Carlo generators~\cite{Fischer:2016zzs,Pirner:2018ccp}, and
inevitably includes additional model-dependent assumptions.

Recently detailed studies of heavy quarkonia~\cite{PSIMULT,Alice:2012Mult}
and open heavy flavor mesons~\cite{Adam:2015ota} in $pp$ collisions
have discovered that similar enhancement with multiplicity also happens
for the production of heavier quarks, charm and bottom. This enhancement
has a quite complicated dependence on the rapidity separation of the
bins used to collect quarkonia and light particles, on the existence
of rapidity gaps between the heavy hadrons and colliding protons~\cite{Siddikov:2020pjh},
as well as (possibly) on the quantum numbers of the produced quarkonia
states~\cite{Siddikov:2020lnq}. While the production of these mesons
can be described in the two-pomeron fusion picture~\cite{Bodwin:1994jh,Maltoni:1997pt,Brambilla:2008zg,Feng:2015cba,Brambilla:2010cs,Baranov:2015laa,Baranov:2016clx,Baier:1981uk,Berger:1980ni,Chang:1979nn,Maciula:2013wg},
as was pointed out in~~\cite{Fischer:2016zzs}, the description
of the multiplicity dependence presents challenges for the established
two-pomeron paradigm. For this reason a number of new mechanisms have
been suggested for its description: \emph{e.g}. the percolation approach~\cite{PER},
a modification of the slope of the elastic amplitude~\cite{Kopeliovich:2013yfa}
or contributions of multipomeron diagrams~\cite{LESI,Siddikov:2019xvf,KMRS,MOSA,Schmidt:2018gep,Levin:2016enb}. 

In view of the similarity of multiplicity enhancements observed in
strange, charm and bottom sectors, it is very desirable to describe
the phenomenon for all flavors in the same framework. Since the strange
quarks have very light mass, in general it is very challenging to
apply the theoretical tools which rely on the perturbative QCD or
heavy quark mass limit for their justification. Nevertheless, in the
kinematics of very large transverse momenta $p_{T}$ of produced strange
hadrons, the latter variable effectively plays the role of hard scale
which partially justifies the use of such perturbative tools.

In what follows we will use the color dipole (CGC/Saturation) approach~\cite{GLR,McLerran:1993ni,McLerran:1993ka,McLerran:1994vd,MUQI,MV,gbw01:1,Kopeliovich:2002yv,Kopeliovich:2001ee}
which was previously applied to $D$- and $B$-meson production in~~\cite{Binnewies:1998vm,Kniehl:1999vf,Ma:2018bax,Goncalves:2017chx,Schmidt:2020fgn}.
The generalization of this framework to high-multiplicity events is
well-known from the literature~\cite{KOLEB,KLN,DKLN,Kharzeev:2000ph,Kovchegov:2000hz,LERE,Lappi:2011gu,Ma:2018bax}
and allows to explain the multiplicity dependence observed in both
the charm and bottom sectors. Our analysis will be mostly focused
on the production of kaons and $\Lambda$-baryons, due to lack of
information about fragmentation functions of other strange hadrons.

The paper is structured as follows. In the next Section~\ref{sec:Evaluation}
we describe a framework for strangeness production in the CGC/Sat
approach. In Section~\ref{sec:Numer} we make numerical estimates
for the cross-sections and compare with available experimental data.
In Section~\ref{subsec:MultiplicityGeneralities} we discuss the
multiplicity dependence of strange hadrons in the large-$p_{T}$ kinematics
and demonstrate that our approach can describe the experimentally
observed dependence for kaons and $\Lambda$-baryons. Finally, in
Section~\ref{sec:Conclusions} we draw conclusions.

\section{Production of strange hadrons via fragmentation}

\label{sec:Evaluation} We assume that all strange hadrons are produced
via a fragmentation mechanism, and we will perform our evaluations
within the framework developed earlier in~\cite{Binnewies:1998vm,Kniehl:1999vf,Ma:2018bax,Goncalves:2017chx,Schmidt:2020fgn}.
In this approach the cross-section is related to the quark pair $\bar{Q}Q$
production cross-section by 
\begin{equation}
\frac{d\sigma_{pp\to M+X}}{dy\,d^{2}p_{T}}=\sum_{i}\int_{x_{Q}}^{1}\frac{dz}{z^{2}}D_{i}\left(\frac{x_{Q}(y)}{z}\right)\,\frac{d\sigma_{pp\to\bar{Q}_{i}Q_{i}+X}}{dy^{*}d^{2}p_{T}^{*}}\label{eq:fragConvolution}
\end{equation}
where $y$ is the rapidity of the produced strange hadron, $y^{*}=y-\ln z$
is the rapidity of the quark, $p_{T}$ is the transverse momentum
of the produced strange hadron, $D_{i}(z)$ is the fragmentation function
which describes the formation of a given final state from a parton
of flavor $i$, and $d\sigma_{pp\to\bar{Q}_{i}Q_{i}+X}/dy^{*}$ is
the cross-section of quark pair production with quark rapidity $y^{*}$
and transverse momentum $p_{T}^{*}=p_{T}/z$. For the fragmentation
functions of kaons and $\Lambda$-baryons we will use the expressions
available from the literature (see the Appendix~\ref{sec:FragFunctions}
for details). Up to the best of our knowledge, currently there is
no data for the fragmentation functions for $\Omega,\,\Xi$-baryons
and $\phi$-mesons, for this reason we will not consider them in what
follows. Naturally, the dominant contribution in the strange sector
stems from the strange quarks, although there are also contributions
from other flavors. In what follows we will focus on the evaluation
of the cross-section $d\sigma_{pp\to\bar{Q}_{i}Q_{i}+X}/dy^{*}d^{2}p_{T}^{*}$
which appears in the integrand of~(\ref{eq:fragConvolution}).

\begin{figure}
\includegraphics[width=9cm]{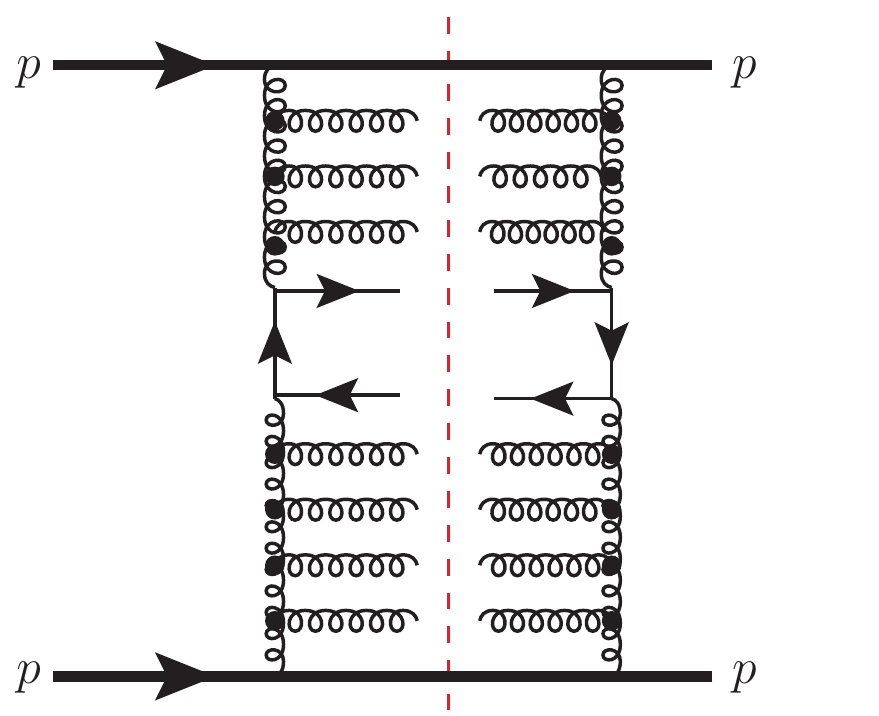}\includegraphics[width=9cm]{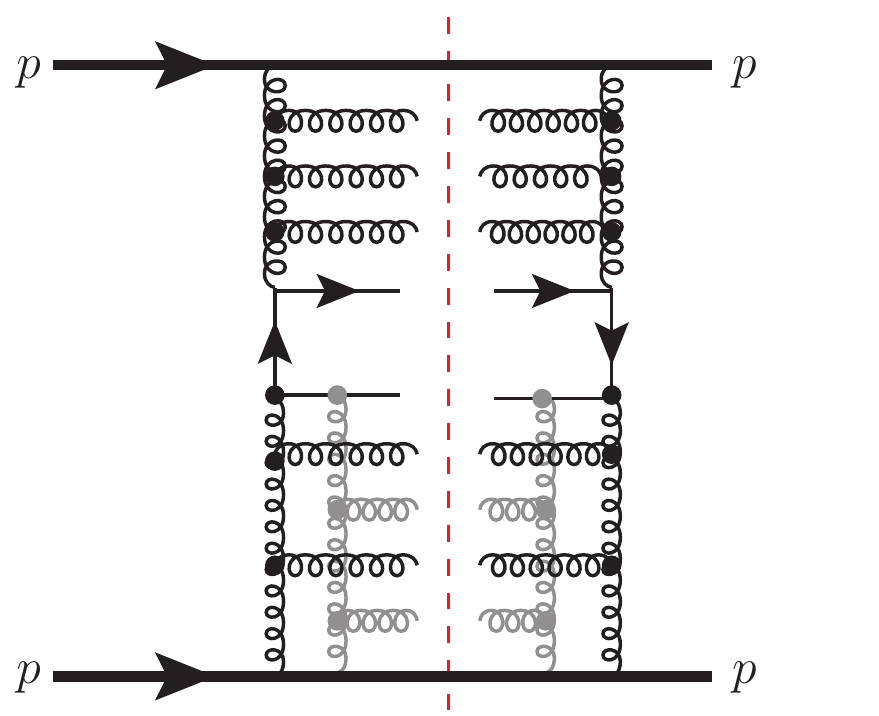}\caption{\label{fig:DipoleCrossSections-23Pom}Left plot: The leading order
two-pomeron mechanism of quark pair production. The diagram includes
two cut pomerons (upper and lower gluon ladders). Right plot: Example
of 3-pomeron mechanisms which might be relevant in the small-$p_{T}$
kinematics (additional reggeons shown with gray color, not considered
in this paper for the reasons discussed in the text). In all plots
the vertical dashed line stands for unitarity cuts. A summation over
all possible permutations of gluon vertices in the quark line (loop)
is implied.}
\end{figure}

In high energy kinematics, the inclusive production gets its dominant
contribution from the fusion of two pomerons, which for heavy quarkonia
production is given by the diagram shown in the left panel of Figure~\ref{fig:DipoleCrossSections-23Pom}.
In the rest frame of one of the protons, this process might be viewed
as a fluctuation of the incoming virtual gluon into a heavy $\bar{Q}Q$
pair, with subsequent scattering of the $\bar{Q}Q$ dipole on the
target proton. In the kinematics of LHC experiments the average light-cone
momentum fractions $x_{1,2}$ carried by gluons are very small ($\ll1$),
and the gluon densities are enhanced. This enhancement implies that
there could be sizable corrections from multiple pomeron exchanges
between the heavy dipole and the target, which are formally suppressed
for small dipoles. For this reason, instead of a hard process on individual
partons it is more appropriate to use the color dipole framework (also
known as CGC/Sat)~\cite{GLR,McLerran:1993ni,McLerran:1993ka,McLerran:1994vd,MUQI,MV,gbw01:1,Kopeliovich:2002yv,Kopeliovich:2001ee}.
At high energies, the color dipoles are eigenstates of interaction,
and therefore they can be used as universal elementary building blocks,
automatically accumulating both the hard and soft fluctuations~\cite{Nikolaev:1994kk}.
In fact, the light-cone color dipole framework has been successfully
applied to phenomenological descriptions of both hadron-hadron and
lepton-hadron collisions~\cite{Kovchegov:1999yj,Kovchegov:2006vj,Balitsky:2008zza,Kovchegov:2012mbw,Balitsky:2001re,Cougoulic:2019aja,Aidala:2020mzt,Ma:2014mri}.
Another advantage of the CGC/Sat framework is that it allows a relatively
straightforward extension for the description of high-multiplicity
events, as discussed in~\cite{KOLEB,KLN,DKLN,Kharzeev:2000ph,Kovchegov:2000hz,LERE,Lappi:2011gu,Ma:2018bax}.

In the dipole approach, the quark production cross-section is given
by~\cite{Ma:2018bax,Goncalves:2017chx} 
\begin{eqnarray}
 &  & \frac{d\sigma_{pp\to\bar{Q}_{i}Q_{i}+X}\left(y,\,\sqrt{s}\right)}{dy\,d^{2}p_{T}}=\,\int d^{2}k_{T}x_{1}\,g\left(x_{1},\,\boldsymbol{p}_{T}-\boldsymbol{k}_{T}\right)\int_{0}^{1}dz\int_{0}^{1}dz'\label{FD1-2}\\
 &  & \times\,\,\,\int\frac{d^{2}r_{1}}{4\pi}\,\int\frac{d^{2}r_{2}}{4\pi}e^{i\left(r_{1}-r_{2}\right)\cdot\boldsymbol{k}_{T}}\,\Psi_{\bar{Q}Q}^{\dagger}\left(r_{2},\,z,\,p_{T}\right)\Psi_{\bar{Q}Q}^{\dagger}\left(r_{1},\,z,\,p_{T}\right)\nonumber \\
 &  & \times N_{M}\left(x_{2}(y);\,\vec{r}_{1},\,\vec{r}_{2}\right),\nonumber \\
 &  & x_{1,2}\approx\frac{\sqrt{m_{M}^{2}+\langle p_{\perp M}^{2}\rangle}}{\sqrt{s}}e^{\pm y}
\end{eqnarray}
where $y$ and $\boldsymbol{p}_{T}$ are the rapidity and transverse
momenta of the produced strange quark in the center-of-mass frame
of the colliding protons; $\boldsymbol{k}_{T}$ is the transverse
momentum of the strange quark with respect to incident gluon; $g\left(x_{1},\,\boldsymbol{p}_{T}\right)$
in the first line of~(\ref{FD1-2}) is the unintegrated gluon PDF;
$\Psi_{g\to\bar{Q}Q}(r,\,z)$ is the light-cone wave function of the
$\bar{Q}Q$ pair with transverse separation between quarks $r$ and
the light-cone fraction carried by the quark $z$. In general this
is a nonperturbative object, and there is no model-independent way
to evaluate it~\footnote{We would like to remind that in case of $D$- and $B$-meson production
studied earlier in~\cite{Binnewies:1998vm,Kniehl:1999vf,Ma:2018bax,Goncalves:2017chx,Schmidt:2020fgn}
the scale in the small-$p_{T}$ kinematics was set by the heavy quark
mass, so the use of the framework was justified up to $p_{T}\approx0$.
In case of the strangeness production this is no longer true, and
we have to restrict our consideration to the large-$p_{T}$ domain
only.}. For this reason, in what follows we will restrict our consideration
to the kinematics of large transverse momenta of produced hadrons,
which in view of~(\ref{FD1-2}) implies that typical sizes of the
dipoles are also small, $\sim1/p_{T}$. In this kinematics we may
use standard perturbative expressions~\cite{Dosch:1996ss,Bjorken:1970ah}
\begin{align}
\Psi_{T}^{\dagger}\left(r_{2},\,z,\,Q^{2}\right)\Psi_{T}\left(r_{1},\,z,\,Q^{2}\right) & =\frac{\alpha_{s}N_{c}}{2\pi^{2}}\left\{ \epsilon_{f}^{2}\,K_{1}\left(\epsilon_{f}r_{1}\right)K_{1}\left(\epsilon_{f}r_{2}\right)\left[e^{i\theta_{12}}\,z^{2}+e^{-i\theta_{12}}(1-z)^{2}\right]\right.\\
 & \left.+m_{f}^{2}K_{0}\left(\epsilon_{f}r_{1}\right)K_{0}\left(\epsilon_{f}r_{2}\right)\right\} ,\nonumber \\
\Psi_{L}^{\dagger}\left(r_{2},\,z,\,Q^{2}\right)\Psi_{L}\left(r_{1},\,z,\,Q^{2}\right) & =\frac{\alpha_{s}N_{c}}{2\pi^{2}}\,\left\{ 4Q^{2}z^{2}(1-z)^{2}K_{0}\left(\epsilon_{f}r_{1}\right)K_{0}\left(\epsilon_{f}r_{2}\right)\right\} ,
\end{align}
\begin{equation}
\epsilon_{f}^{2}=z\,(1-z)\,Q^{2}+m_{f}^{2}
\end{equation}
\begin{equation}
\left|\Psi^{(f)}\left(r,\,z,\,Q^{2}\right)\right|^{2}=\left|\Psi_{T}^{(f)}\left(r,\,z,\,Q^{2}\right)\right|^{2}+\left|\Psi_{L}^{(f)}\left(r,\,z,\,Q^{2}\right)\right|^{2}
\end{equation}
The meson production amplitude $N_{M}$ depends on the mechanism of
$Q\bar{Q}$ pair formation. For the case of the two-pomeron fusion,
in leading order it is given by~\cite{Goncalves:2017chx,Schmidt:2020fgn}
\begin{eqnarray}
N_{M}\left(x,\,\,\vec{r}_{1},\,\vec{r}_{2}\right) & = & -\frac{1}{2}N\left(x,\,\vec{r}_{1}-\vec{r}_{2}\right)-\frac{1}{16}\left[N\left(x,\,\vec{r}_{1}\right)+N\left(x,\,\vec{r}_{2}\right)\right]-\frac{9}{8}N\left(x,\,\bar{z}\left(\vec{r}_{1}-\vec{r}_{2}\right)\right)\label{eq:N2}\\
 &  & +\frac{9}{16}\left[N\left(x,\,\bar{z}\vec{r}_{1}-\vec{r}_{2}\right)+N\left(x,\,\bar{z}\vec{r}_{2}-\vec{r}_{1}\right)+N\left(x,\,\bar{z}\vec{r}_{1}\right)+N\left(x,\,\bar{z}\vec{r}_{2}\right)\right],\nonumber 
\end{eqnarray}
where $N\left(x,\,\vec{\boldsymbol{r}}\right)$ is the color singlet
dipole scattering amplitude. In the LHC kinematics at large transverse
momenta (our principal interest) the natural choice of the saturation
scale is $\mu_{F}\sim p_{T}$, which significantly exceeds the saturation
scale $Q_{s}(x)$. This finding justifies the use of two-pomeron approximation.
However, in the kinematics of smaller $p_{T}$, potentially there
could be multipomeron contributions, like those shown in the right
panel of Figure~\ref{fig:DipoleCrossSections-23Pom}. We will not
consider such contributions since, as we mentioned earlier, we can't
describe the small-$p_{T}$ region, due to lack of the nonperturbative
photon wave function $\Psi_{\bar{Q}Q}$.

The unintegrated gluon PDF, which appears in the prefactor of~(\ref{FD1-2}),
can be related to the integrated PDF $xG\left(x,\,\mu_{F}\right)$
as~\cite{Kimber:2001sc} 
\begin{equation}
x\,g\left(x,\,k^{2}\right)=\left.\frac{\partial\,}{\partial\mu_{F}^{2}}xG\left(x,\,\mu_{F}\right)\right|_{\mu_{F}^{2}=k^{2}},
\end{equation}
and the latter is closely related to the dipole scattering amplitude
$N\left(y,r\right)=\int d^{2}b\,N\left(y,r,b\right)$ via a set of identities~\cite{KOLEB,THOR}
\begin{equation}
\frac{C_{F}}{2\pi^{2}\bar{\alpha}_{S}}N\left(y,\,\vec{r}\right)=\int\frac{d^{2}k_{T}}{k_{T}^{4}}\phi\left(y,k_{T}\right)\,\Bigg(1-e^{i\vec{k}_{T}\cdot\vec{r}}\Bigg);~~~~x\,G\left(x,\,\mu_{F}\right)=\int_{0}^{\mu_{F}}\frac{d^{2}k_{T}}{k_{T}^{2}}\phi\left(x,\,k_{T}\right),\label{GN1-1}
\end{equation}
where $y=\ln(1/x)$. The Eq. (\ref{GN1-1}) can be inverted and gives
the gluon uPDF in terms of the dipole amplitude, 
\begin{equation}
xG\left(x,\,\mu_{F}\right)\,\,=\,\,\frac{C_{F}\mu_{F}}{2\pi^{2}\bar{\alpha}_{S}}\int d^{2}r\,\frac{J_{1}\left(r\,\mu_{F}\right)}{r}\nabla_{r}^{2}N\left(y,\,\vec{r}\right).\label{GN2-1}
\end{equation}
This allows to rewrite the result in a symmetric and self-consistent
form, which allows straightforward generalization for high-multiplicity
events.

\section{Numerical results}

\label{sec:Numer} For the sake of definiteness, in our numerical
evaluations we will take the ``CGC'' parametrization of the dipole
cross-section~\cite{Kowalski:2003hm,Kowalski:2006hc,RESH}, 
\begin{align}
N\left(x,\,\vec{\boldsymbol{r}}\right) & =\sigma_{0}\times\left\{ \begin{array}{cc}
N_{0}\,\left(\frac{r\,Q_{s}(x)}{2}\right)^{2\gamma_{{\rm eff}}(r)}, & r\,\le\frac{2}{Q_{s}(x)}\\
1-\exp\left(-\mathcal{A}\,\ln\left(\mathcal{B}r\,Q_{s}\right)\right), & r\,>\frac{2}{Q_{s}(x)}
\end{array}\right.~,\label{eq:CGCDipoleParametrization}\\
 & \mathcal{A}=-\frac{N_{0}^{2}\gamma_{s}^{2}}{\left(1-N_{0}\right)^{2}\ln\left(1-N_{0}\right)},\quad\mathcal{B}=\frac{1}{2}\left(1-N_{0}\right)^{-\frac{1-N_{0}}{N_{0}\gamma_{s}}},\\
 & Q_{s}(x)=\left(\frac{x_{0}}{x}\right)^{\lambda/2},\,\,\gamma_{{\rm eff}}(r)=\gamma_{s}+\frac{1}{\kappa\lambda Y}\ln\left(\frac{2}{r\,Q_{s}(x)}\right),\label{eq:gamma}\\
 & \gamma_{s}=0.762,\quad\lambda=0.2319,\quad\sigma_{0}=21.85\,{\rm mb},\quad x_{0}=6.2\times10^{-5}\label{eq:gamma2}
\end{align}
which is widely used in the literature, and in what follows we will
also focus on $K_{S}^{0}$-meson production. The fragmentation functions
for $K^{\pm}$ and $K_{S}^{0}$ are constrained by the isospin symmetry
relation~\cite{Albino:2008fy} 
\begin{equation}
D_{i}^{K_{S}^{0}}\left(z,\,\mu^{2}\right)=\frac{1}{2}D_{\bar{i}}^{K^{\pm}}\left(z,\,\mu^{2}\right),
\end{equation}
and for this reason the cross-sections for $K_{S}^{0}$ and $K^{\pm}$
production are proportional to each other with very good precision.
For $\Lambda$-baryons we found only one parametrization for the fragmentation
function only~\cite{Albino:2008fy}, and as shown in Appendix~(\ref{sec:FragFunctions}),
at large-$z$ its strange flavor component is approximately proportional
to that of kaons; therefore the $\Lambda$-baryon cross-sections are
proportional to that of kaons in the large-$p_{T}$ kinematics.

\begin{figure}
\includegraphics[width=16cm]{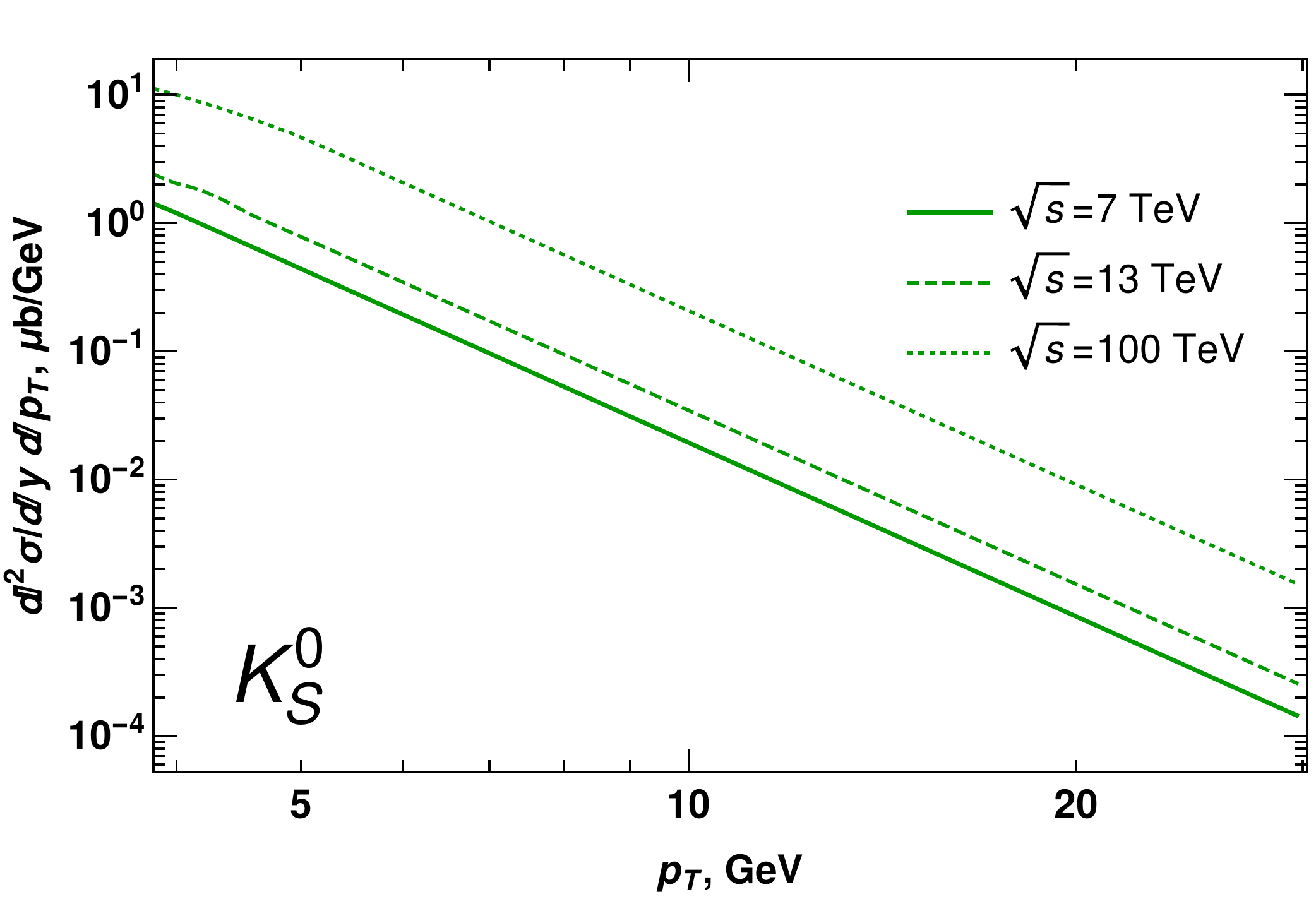}

\caption{\label{fig:K0S_pTDependence}The $p_{T}$-dependence of the cross-section
$d\sigma/dp_{T}$ for $K_{S}^{0}$-meson production at central rapidities,
evaluated with the two-pomeron fusion mechanism. Results for $\Lambda$
have similar shape, due to the similarity of their fragmentation functions
(see details in Appendix~\ref{sec:FragFunctions}).}
\label{Diags_DMesons-1} 
\end{figure}

\begin{figure}
\includegraphics[width=9cm]{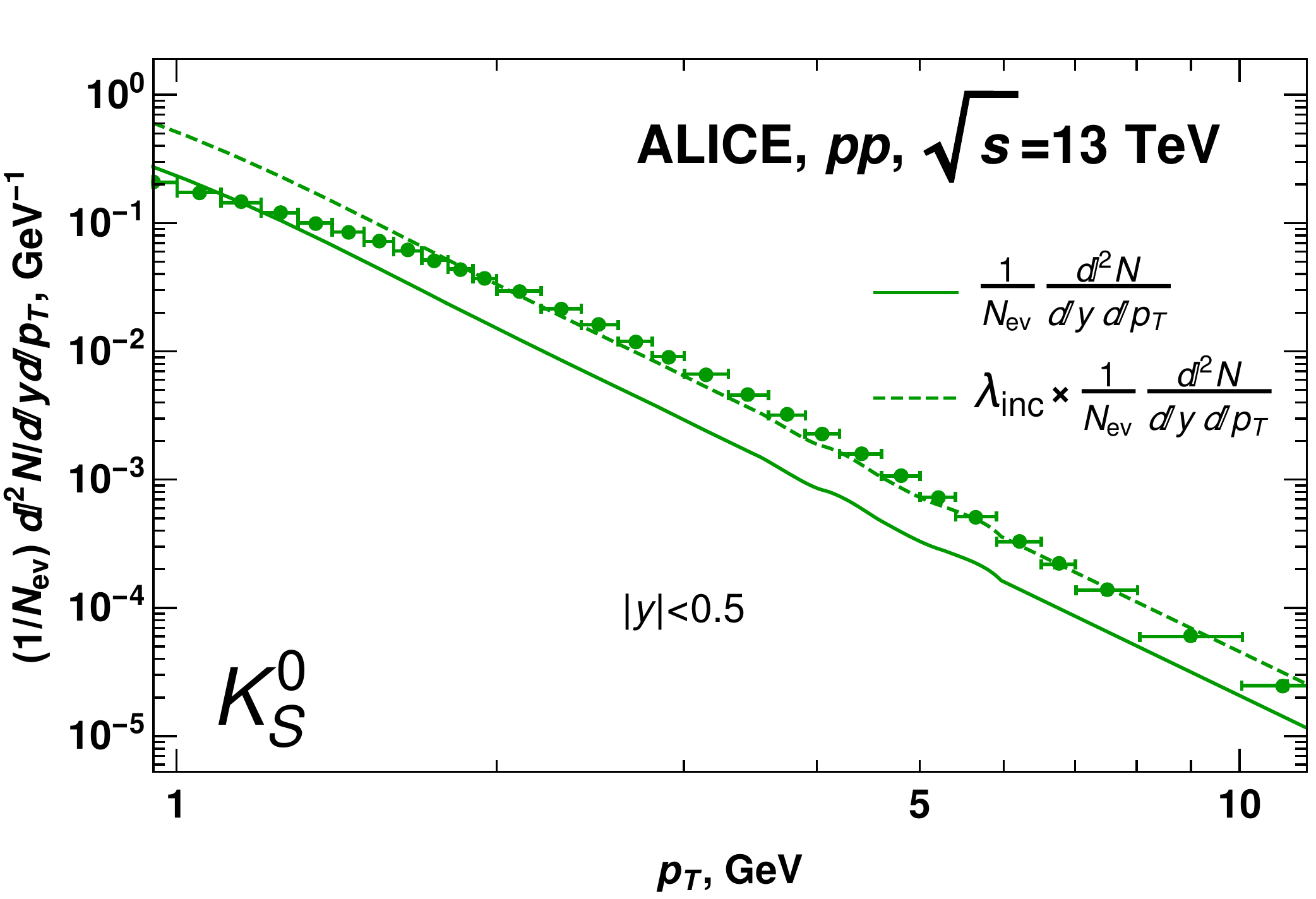}\includegraphics[width=9cm]{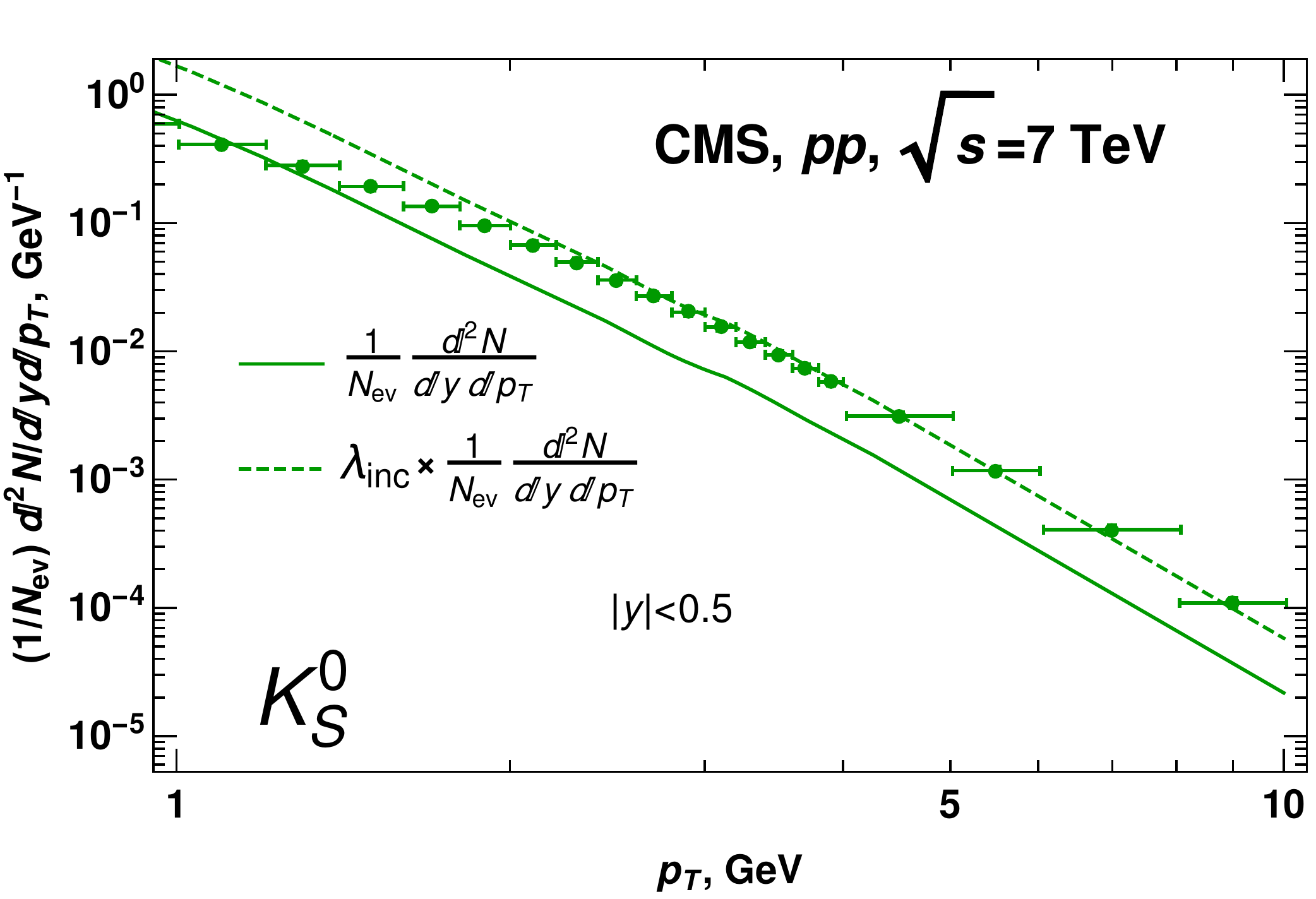}

\includegraphics[width=9cm]{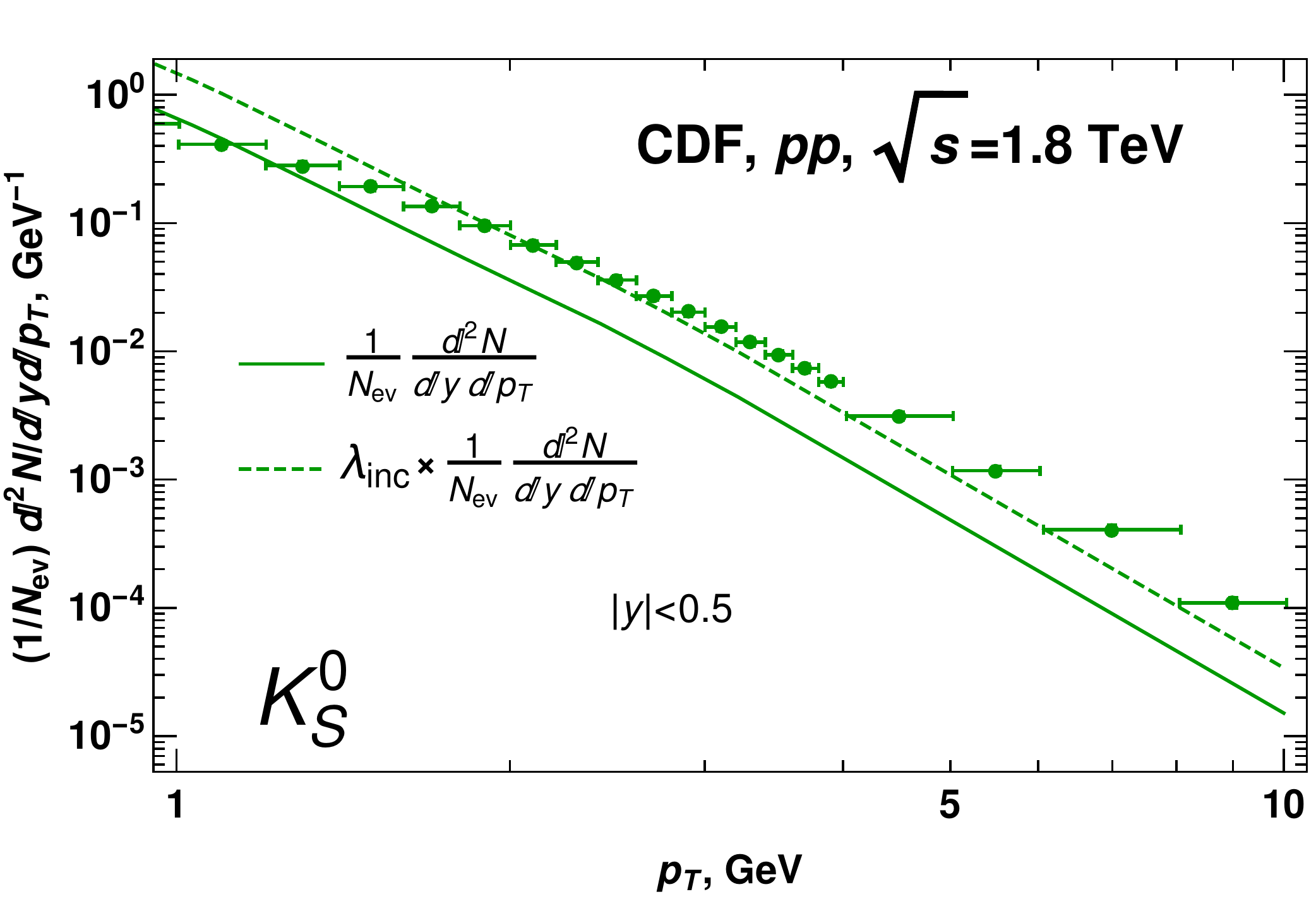}\includegraphics[width=9cm]{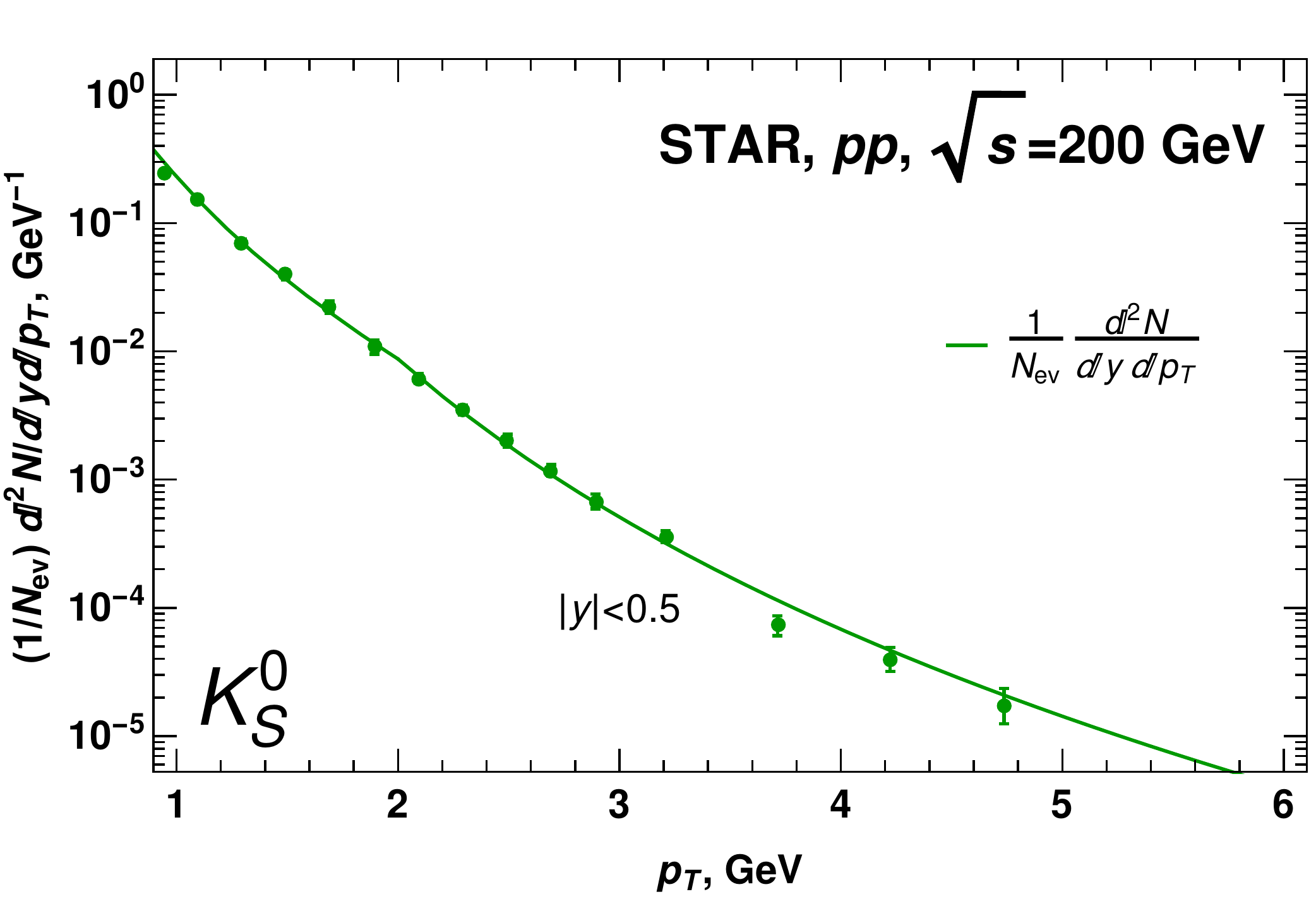}

\caption{\label{fig:K0S_pTDependence-1} \emph{Self-normalized} $p_{T}$-dependence
of the $K_{S}^{0}$-meson yields at central rapidities. Theoretical
predictions (solid line) are compared with experimental data from
ALICE~\cite{Acharya:2019kyh}, CMS~\cite{Khachatryan:2011tm}, CDF~\cite{Acosta:2005pk}
and STAR~\cite{Abelev:2006cs} collaborations. As we explained in
the text, our approach is not very reliable at small $p_{T}$, and
thus the evaluation of the global normalization factor $N_{{\rm ev}}$
might have large nonperturbative corrections. In order to demonstrate
that the description of the \emph{shape} is correct, we have plotted
also the yields multiplied by a constant factor $\lambda_{{\rm inc}}\approx1.7-2.2$
(dashed lines).}
\label{Diags_DMesons-1-1} 
\end{figure}

In the left panel of Figure~\ref{fig:K0S_pTDependence} we show the
results for $K_{S}^{0}$ production cross-section, in the kinematics
of ongoing and planned experiments. As can be seen in the literature,
all the experimental papers contain results for the self-normalized
yields $N_{{\rm ev}}^{-1}\,dN/dy\,dp_{T},$ instead of cross-sections.
The normalization parameter $N_{{\rm ev}}$ is chosen as the total
number of events $N_{{\rm ev}}\sim\int dp_{T}dN/dp_{T}$ or the number
of non-single-diffractive events $N_{{\rm NSD}}$. The self-normalized
yields are difficult to describe in our approach, because the normalization
coefficient gets its dominant contribution from the small-$p_{T}$
region and thus cannot be evaluated reliably. This introduces a normalization
uncertainty in our evaluations of such self-normalized yields. In
Figure~\ref{fig:K0S_pTDependence-1} we show the calculated yields
in comparison with available experimental data from ALICE~\cite{Acharya:2019kyh},
CMS~\cite{Khachatryan:2011tm}, CDF~\cite{Acosta:2005pk} and STAR~\cite{Abelev:2006cs}
collaborations~\footnote{We also would like to mention that the CMS and STAR data were not
normalized to unity. For example, the CMS data instead of $N_{{\rm ev}}$
used the number of non-single-diffractive events $N_{{\rm NSD}}$,
so we also corrected the normalization of experimental data.}. The STAR data~\cite{Abelev:2006cs} were included by authors of~\cite{Albino:2008fy}
in their global fit of fragmentation functions, for this reason the
description of these data is nearly perfect. For data from LHC and
Tevatron the model provides a very reasonable description of the shape,
although, as expected, there is a mismatch in the normalization by
a factor of two. The theoretical shape of the $p_{T}$-dependence
starts deviating from the experimental data in the region $p_{T}\lesssim$2~GeV,
where nonperturbative effects become pronounced.

\section{Multiplicity dependence}

\label{subsec:MultiplicityGeneralities}Since the dipole approach~(\ref{FD1-2},
\ref{eq:N2},~\ref{eq:CGCDipoleParametrization}) provides a reasonable
description of the strangeness production in the large-$p_{T}$ kinematics,
we can apply it to the study of the dependence on the number of charged
particles $N_{{\rm ch}}$ co-produced together with a given strange
hadron. The extension of the color dipole (CGC/Sat) framework to the
description of high-multiplicity events is quite straightforward,
as was discussed in~\cite{KOLEB,KLN,DKLN,Kharzeev:2000ph,Kovchegov:2000hz,LERE,Lappi:2011gu,Ma:2018bax,Schmidt:2020fgn,Siddikov:2019xvf,Siddikov:2020pjh}.
In what follows we will briefly summarize the main results which will
be used for the phenomenological estimates of the multiplicity dependence.
In view of the Local Parton Hadron Duality (LPHD) hypothesis~\cite{LPHD1,LPHD2,LPHD3},
the multiplicity of produced hadrons in a given event is directly
proportional to the number of partons produced in a collision. For
this reason, the study of high multiplicity events in different channels
allows to understand better the onset of the saturation regime in
high energy collisions.

The probability of multiplicity fluctuations decreases rapidly as
function of the number of produced charged particles $N_{{\rm ch}}$~\cite{Abelev:2012rz},
for this reason in the study of the multiplicity dependence it is
more common to use a self-normalized ratio~\cite{Acharya:2019kyh}
\begin{eqnarray}
\frac{dN_{M}/dy}{\langle dN_{M}/dy\rangle}\,\, & = & \frac{d\sigma_{M}\left(y,\,\eta,\,\sqrt{s},\,n\right)/dy}{d\sigma_{M}\left(y,\,\eta,\,\sqrt{s},\,\langle n\rangle=1\right)/dy}/\frac{d\sigma_{{\rm ch}}\left(\eta,\,\sqrt{s},\,Q^{2},\,n\right)/d\eta}{d\sigma_{{\rm ch}}\left(\eta,\,\sqrt{s},\,Q^{2},\,\langle n\rangle=1\right)/d\eta}\label{eq:NDef}
\end{eqnarray}
where $n=N_{{\rm ch}}/\langle N_{{\rm ch}}\rangle$ is the relative
enhancement of the number of charged particles in a given observation
window (e.g. pseudorapidity bin $(\eta-\Delta\eta/2,\,\,\eta+\Delta\eta/2)$);
$d\sigma_{M}(y,\,\sqrt{s},\,n)$ is the strange hadron $M$ production
cross-section, with rapidity $y$ and $N_{{\rm ch}}=n\,\langle N_{{\rm ch}}\rangle$
charged particles; $d\sigma_{{\rm ch}}(y,\,\sqrt{s},\,n)$ is the
total production cross-section for $N_{{\rm ch}}=n\,\langle N_{{\rm ch}}\rangle$
charged particles in the same observation window. Since the cross-sections
are proportional to the probability to produce a given final state,
the ratio~(\ref{eq:NDef}) might be interpreted as a \emph{conditional}
probability to produce a strange hadron $M$ in a $pp$ collision
in which $N_{{\rm ch}}$ charged particles are produced.

We expect that even in high multiplicity events each gluon cascade
(``pomeron'') should satisfy the nonlinear Balitsky-Kovchegov equation,
and therefore the dipole amplitude~(\ref{eq:CGCDipoleParametrization})
should keep its form, although the value of the saturation scale $Q_{s}$
might be modified. As was demonstrated in~\cite{KOLEB,KLN,DKLN},
the observed number of charged multiplicity $dN_{{\rm ch}}/dy$ of
soft hadrons in $pp$ collisions is proportional to the saturation
scale $Q_{s}^{2}$ (modulo logarithmic corrections), and therefore
in the dipole framework the events with large multiplicity might be
described by simply rescaling $Q_{s}^{2}$ as a function of $n$~\cite{KOLEB,KLN,DKLN,Kharzeev:2000ph,Kovchegov:2000hz,LERE,Lappi:2011gu},

\begin{equation}
Q_{s}^{2}\left(x,\,b;\,n\right)\approx n\,Q^{2}\left(x,\,b\right).\label{QSN-1}
\end{equation}
The accuracy of the approximation~(\ref{QSN-1}) was tested in~\cite{Ma:2018bax},
and it was found that its error does not exceed 10 per cent in the
region of interest ($n\lesssim10$), on par with the precision of
current evaluations. Therefore, in what follows we will use~(\ref{QSN-1})
for our estimates. While at LHC energies it is expected that the typical
values of the saturation scale $Q_{s}\left(x,\,b\right)$ fall into
the range 0.5-1 ${\rm GeV}$, from~(\ref{QSN-1}), we can see that
in events with enhanced multiplicity this parameter might lead to
an interplay of the large-$Q_{s}$ and large-$p_{T}$ limits. Since
increasing multiplicity and increasing energy (decreasing $x$) affect
$Q_{s}^{2}$ in a similar way, the study of the high-multiplicity
events allows to study a deeply saturated regime, which determines
the dynamics of all processes at significantly higher energies.

\begin{figure*}
\includegraphics[width=9cm]{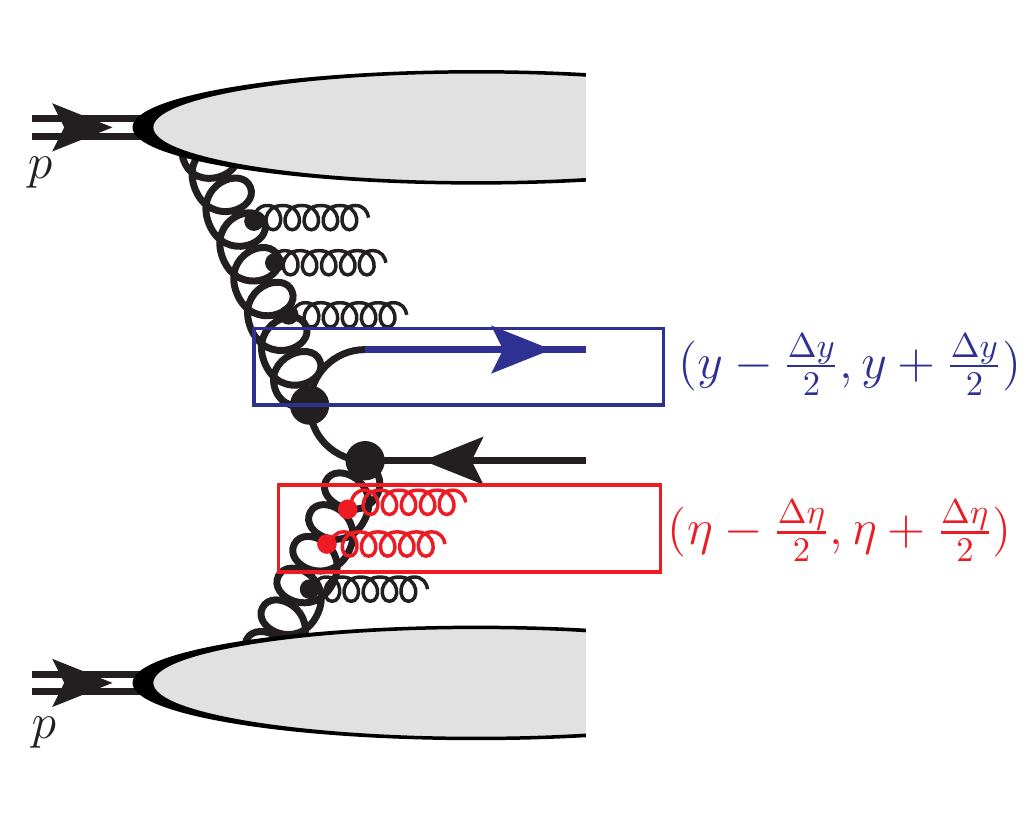}\includegraphics[width=9cm]{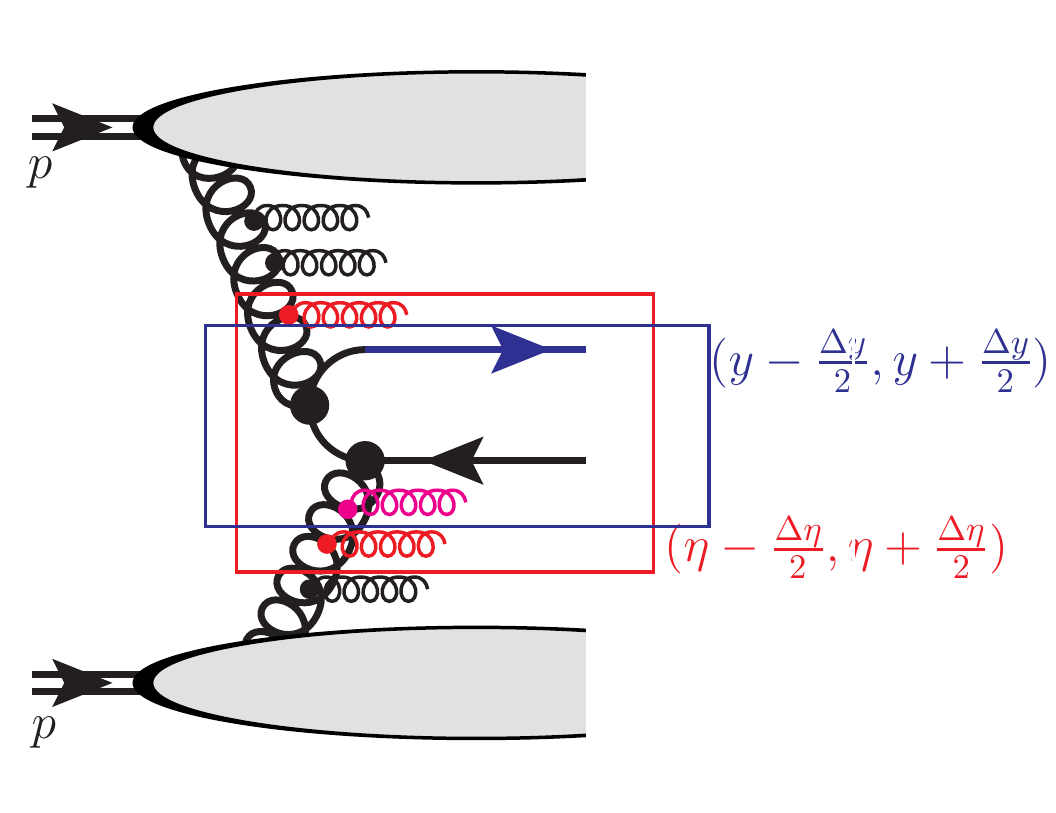}

\caption{(color online) Demonstration that sharing of the enhanced multiplicity
of the whole process between individual pomerons depends on the experimental
setup. Left plot: The experimental setup in which the rapidity bin
used for collection of strange hadrons (blue box) does not overlap
with the bin used for the collection of charged particles (red box).
The elevated multiplicity in this case should be unambiguously attributed
to the lower pomeron. Right plot: The experimental setup when the
bins partially overlap. For the partons in the intersection region
(magenta color) the assignment to upper or lower pomerons depends
on the position of large-$p_{T}$ quark inside the bin~($y-\Delta y/2,\,y+\Delta y/2$).
In the final result we should average over all possible rapidities
of strange quark inside the bin. \label{fig:DiagMultiplicityDistribution}}
\end{figure*}

Since at high energies each pomeron hadronizes independently~\cite{Korchemsky:2001nx},
the observed enhancement of multiplicity in the whole process must
be shared between all pomerons which might contribute in a given rapidity
window. For this reason, for phenomenological estimates of the multiplicity
dependence it is important if the rapidity bin used to collect charged
particles $N_{{\rm ch}}$ overlaps with the bin used for the observation
of strange large-$p_{T}$ hadrons, as explained in Figure~\ref{fig:DiagMultiplicityDistribution}.

In the limit of large-$p_{T}$ the typical sizes of the dipoles are
small, $r\sim1/p_{T}$, so we may expect from~(\ref{FD1-2},~\ref{eq:CGCDipoleParametrization},~\ref{QSN-1})
that the contribution of each cut pomeron to the multiplicity dependence
is given by the factor$\sim n_{i}^{\langle\gamma_{{\rm eff}}\rangle}$,
where $n_{i}$ is the relative enhancement of multiplicity assigned
to a given pomeron, and the value of the parameter $\gamma_{{\rm eff}}$
is given in~(\ref{eq:gamma},~\ref{eq:gamma2}). For the configuration
when the bins used to collect charged and strange particles are separated
by rapidity (see the left panel of the Figure~\ref{fig:DiagMultiplicityDistribution})
the multiplicity is assigned to one of the pomerons, so the expected
multiplicity dependence of the cross-section is $\sim n^{\langle\gamma_{{\rm eff}}\rangle}$.
For the case when the strange and charged particle bins partially
overlap (as shown in the right panel of the Figure~\ref{fig:DiagMultiplicityDistribution}),
we should average over all possible partitions of the observed number
of charged particles. This evaluation technically is quite complicated,
although we can assume with good precision that the multiplicity enhancement
is shared equally between both pomerons~\cite{LESI}. For this reason
in this case we expect that the multiplicity cross-section would be
$\sim(n/2)^{2\langle\gamma_{{\rm eff}}\rangle}$.

\begin{figure}
\includegraphics[width=18cm]{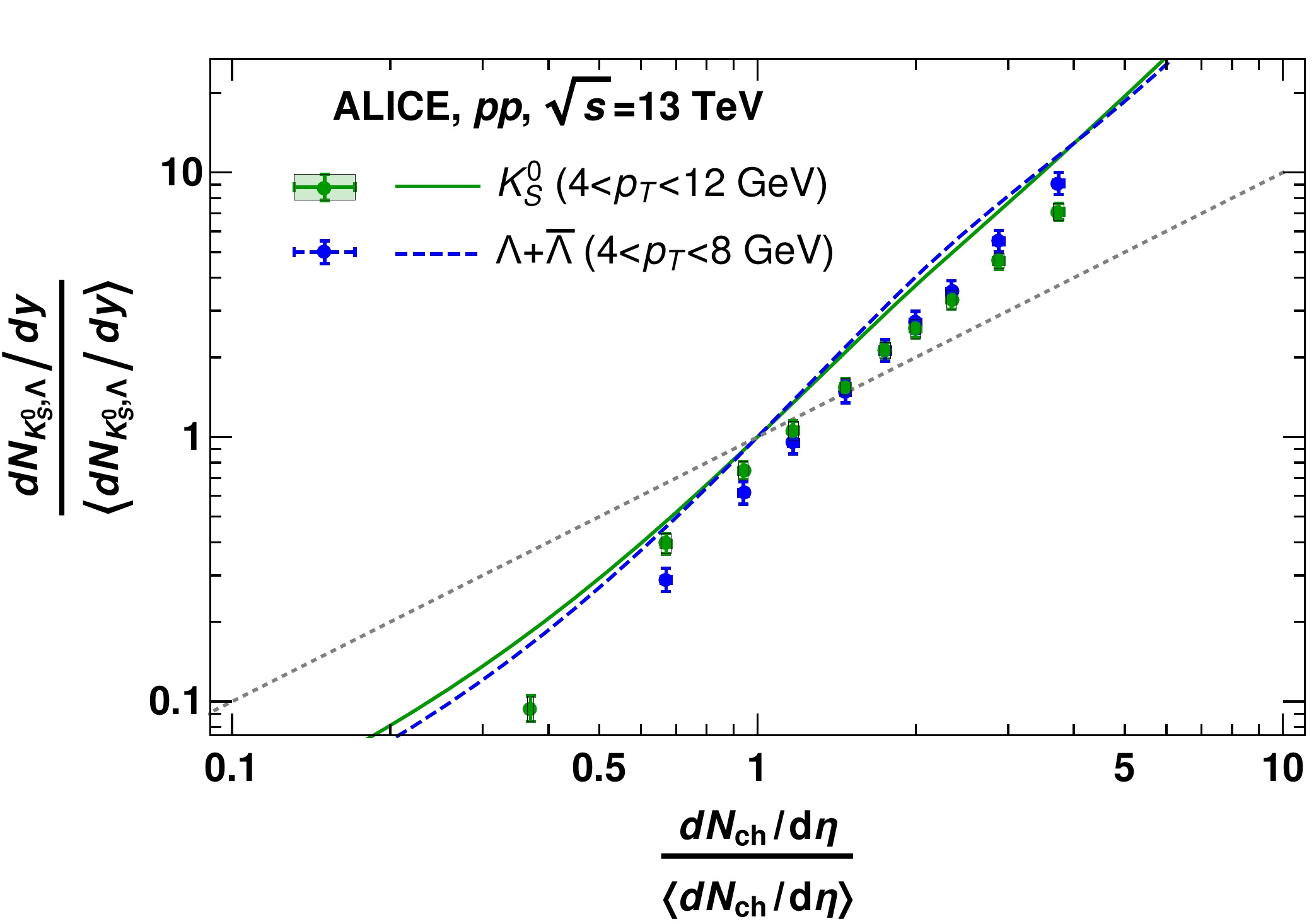}

\caption{\label{DiagsMultiplicityStrange}Comparison of the theoretical multiplicity
dependence for $K_{S}^{0}$ meson (production solid curve) and $\Lambda$
baryons (dashed curve) with experimental data from ALICE~\cite{Acharya:2019kyh}.
For the sake of reference we have also shown a dotted line, which
corresponds to a linear dependence. The charged particles and strange
hadrons are collected at central rapidities.}
\end{figure}

Currently the data on multiplicity dependence of strange hadrons are
available from the ALICE experiment~\cite{Acharya:2019kyh}. As we
can see from Figure~\ref{DiagsMultiplicityStrange}, the theoretical
curves can describe reasonably well the slope of the experimentally
observed $n$-dependence (in logarithmic coordinates), although apparently
overestimate all the experimental points by the same normalization
factor $\sim1.2$. We would like to stress that by definition at the
point $n=1$ the self-normalized ratio~(\ref{eq:NDef}) equals one.
This condition is fulfilled in our theoretical curves, and therefore
we believe that the normalization of our curves is correct.

\section{Conclusions}

\label{sec:Conclusions}In this paper we studied the production of
strange hadrons in the color dipole approach. We found that the CGC/Sat
approach can describe the \emph{shapes} of $p_{T}$ distributions
in the large-$p_{T}$ kinematics, although it might not be very reliable
for smaller $p_{T}$. The latter restriction implies that the suggested
approach cannot be applied to $p_{T}$-integrated observables, which
get its dominant contribution from the nonperturbative small-$p_{T}$
region. As a consequence, our predictions for experimentally measurable
self-normalized yields suffer from a global normalization uncertainty,
which complicates the direct comparison of model predictions with
data, even for large $p_{T}$. For this reason we call experimentalists
to publish also the cross-sections, for which there is unambiguous
separation of small-$p_{T}$ and large-$p_{T}$ physics. We made predictions
for such cross-sections in the kinematics of ongoing and future experiments,
and can provide further predictions on demand.

We also applied the CGC/Sat approach to the description of the multiplicity
dependence measured by ALICE~\cite{Acharya:2019kyh}. Fortunately,
all kaons and $\Lambda$-baryons were collected with sufficiently
large transverse momenta $p_{T}\gtrsim4$ GeV, where our approach
is well justified. We found that the theoretical predictions are in
reasonable agreement with experimental data. Our evaluation is largely
parameter-free and relies only on the choice of the parametrization
for the dipole cross-section~(\ref{eq:CGCDipoleParametrization})
and fragmentation functions of strange hadrons.

This study complements our previous analysis of the multiplicity dependence
of heavier charm and bottom production~\cite{Schmidt:2020fgn} and
demonstrates that at sufficiently large $p_{T}$ it is possible to
describe all of them within the same framework.

\section*{Acknowledgements}

We thank our colleagues at UTFSM University for encouraging discussions.
This research was partially supported by Proyecto Basal FB 0821(Chile)
and Fondecyt (Chile) grant 1180232. Also, we thank Yuri Ivanov for
technical support of the USM HPC cluster, where part of the evaluations
were performed.

\appendix

\section{Fragmentation functions}

\label{sec:FragFunctions} In this section we would like to summarize
briefly the fragmentation functions used in our evaluations. These
functions are nonperturbative objects, which cannot be evaluated from
first principles. For this reason currently their parametrization
is extracted from the phenomenological fits of experimental data.
For the sake of definiteness, for our evaluations we used the fragmentation
functions for kaons and $\Lambda$ from~\cite{Albino:2008fy} (so-called
AKK08 parametrization). The fragmentation functions for $K^{\pm}$
and $K_{S}^{0}$ are constrained by the isospin symmetry relation
\begin{equation}
D_{i}^{K_{S}^{0}}\left(z,\,\mu^{2}\right)=\frac{1}{2}D_{\bar{i}}^{K^{\pm}}\left(z,\,\mu^{2}\right),
\end{equation}
therefore in what follows we will consider only the fragmentation
function of neutral kaons $K_{S}^{0}$. For kaons we checked that
the alternative parametrizations of fragmentation functions DSS17~\cite{deFlorian:2017lwf},
NNPDF~\cite{Bertone:2017tyb} and JAM~\cite{Sato:2016wqj} give
similar results in the region of interest. We have not found parametrizations
for fragmentation functions of strange baryons $\Omega,\,\Xi$, and
neither of $K_{S}^{0*}$ and $\phi$-mesons, and for this reason we
do not consider them in this paper.

\begin{figure}
\includegraphics[height=6.5cm]{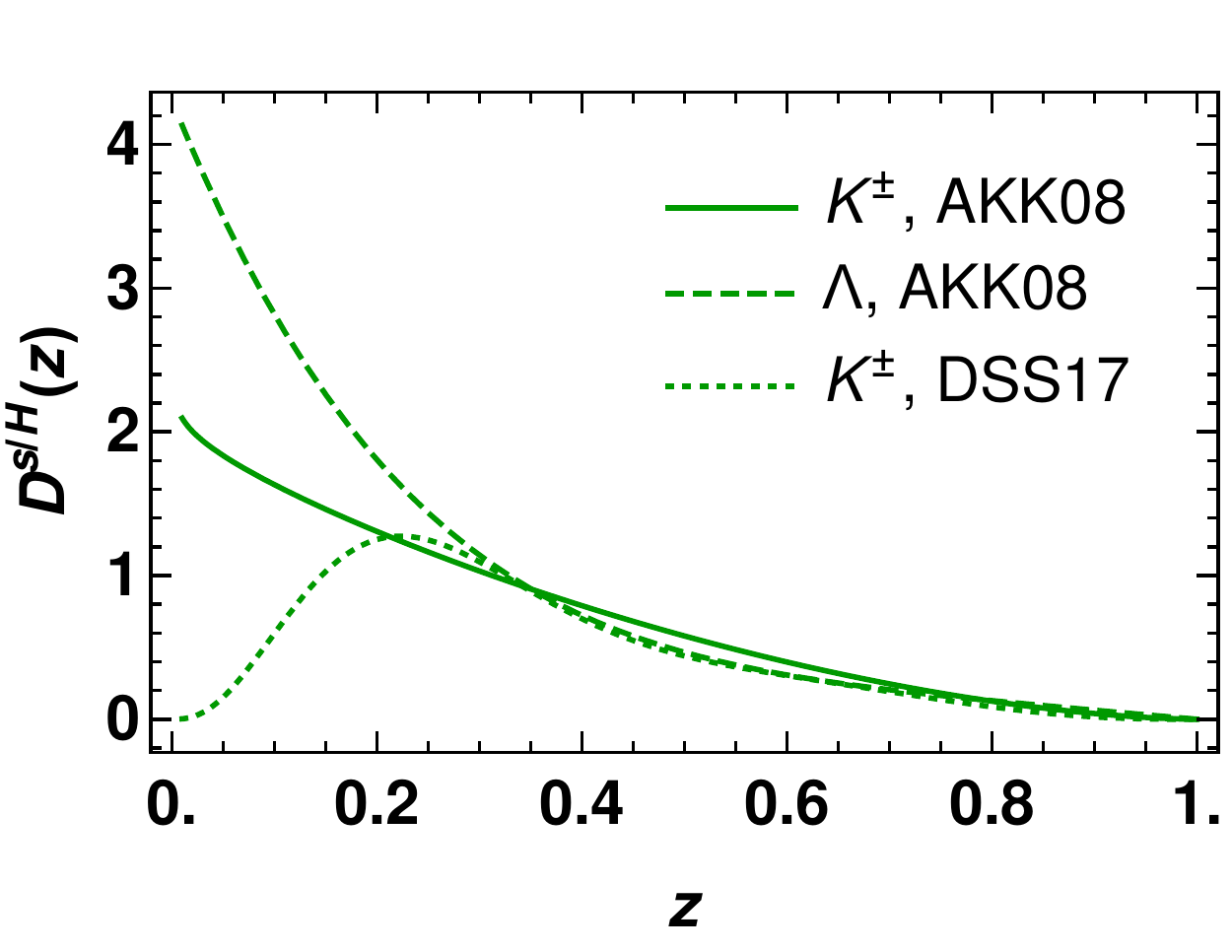}\includegraphics[height=6.5cm]{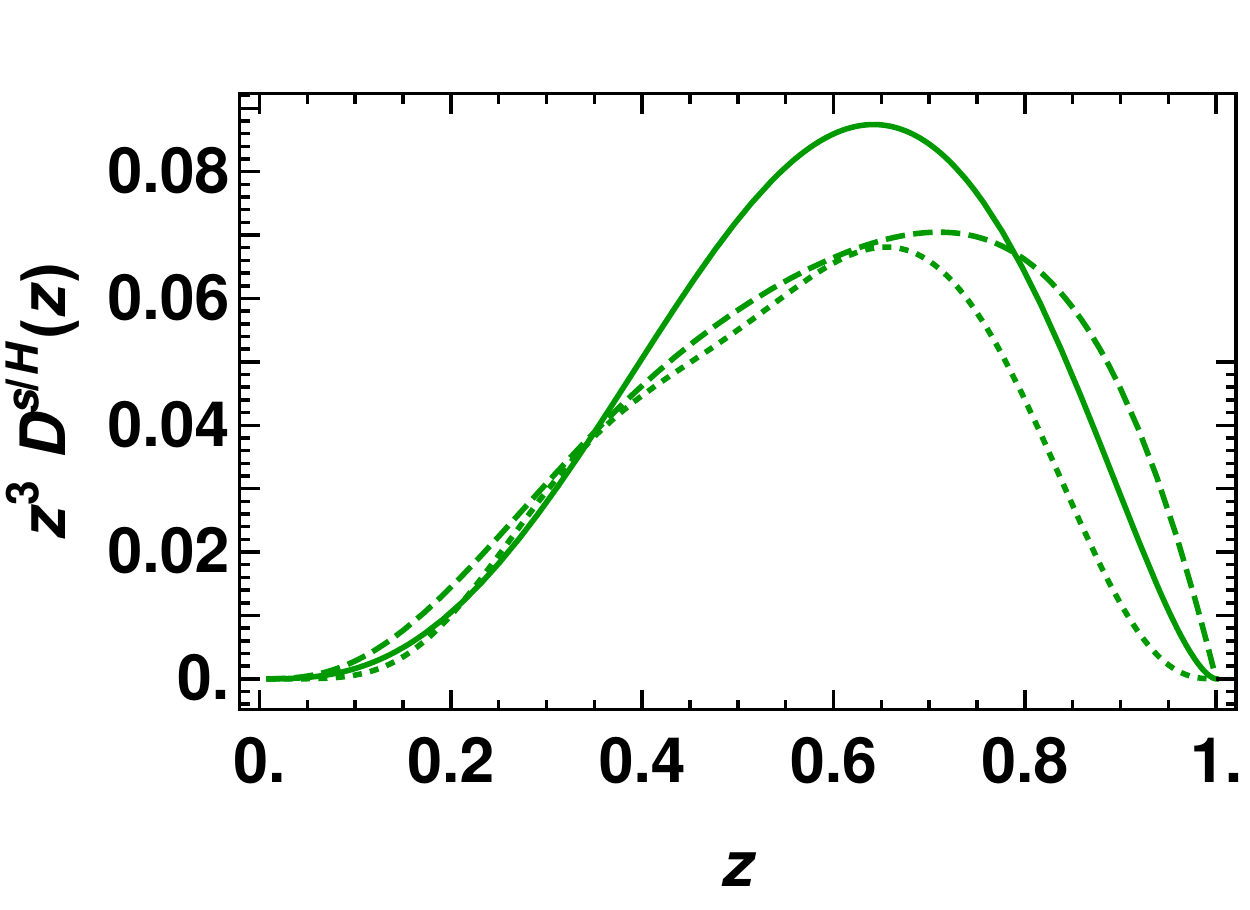}

\caption{\label{fig:fragFunction} Left: The fragmentation function $D^{s/H}$
of kaons and $\Lambda$-baryon ($s$-quark component), evaluated in
AKK08~\cite{Albino:2008fy} and DSS17~\cite{deFlorian:2017lwf}
parametrizations. Right: The same function multiplied by $z^{3}$.
As explained in the text, for the physical cross-section in the large-$p_{T}$
kinematics the fragmentation function contributes multiplied by an
additional factor $\sim z^{3}$, so the difference of the fragmentation
functions in the small-$z$ domain has a minor effect on physical
observables.}
\label{DiagsMultiplicity1-1} 
\end{figure}

In the AKK08 parametrization~\cite{Albino:2008fy} it is assumed
that the fragmentation function is given by 
\begin{equation}
D^{i/H}(z)=N_{i}z^{a_{i}}(1-z)^{b_{i}}\left[1+c_{i}(1-z)^{d_{i}}\right],
\end{equation}
where $N_{i},a_{i},b_{i},c_{i},d_{i}$ are some numerical coefficients
which depend on the hadron and quark flavor $i$. We expect that for
strange hadrons the largest contribution comes from the fragmentation
of the strange quark, thus we will discuss below the fragmentation
function $D^{s/H}$. As we can see from Figure~\ref{fig:fragFunction},
the parametrizations for $K^{\pm}$ mesons and $\Lambda$-baryons
differ quite substantially in the region of small $z\lesssim0.3$,
although become comparable for all hadrons at larger values of $z$.

In this paper we are mostly interested in the large-$p_{T}$ kinematics,
and it is possible to show that this region has stronger sensitivity
to the region of large $z$. Indeed, as we can see from the structure
of~(\ref{FD1-2}), at large $p_{T}$ the cross-section $d\sigma_{\bar{Q}Q}/dp_{T}^{\bar{Q}Q}$
is suppressed as $\sim\left(1/p_{T}^{\bar{Q}Q}\right)^{n}$ with $n\gtrsim5$.
The momentum of the quark pair $p_{T}^{\bar{Q}Q}$ is related to the
momentum of the strange hadron as $p_{T}^{\bar{Q}Q}=p_{T}/z$, so
this implies that in the integral over the fragmentation fraction
$z$ in (\ref{eq:fragConvolution}) effectively we get an additional
prefactor $\sim z^{n-2}$, which suppresses the contribution of the
small-$z$ domain. As we can see from the right panel of Figure~\ref{fig:fragFunction},
the dominant contribution comes from the region $z\sim0.6-0.8$, where
the difference between fragmentation functions does not exceed a factor
of two.

\end{document}